\documentclass[conference]{IEEEtran}
\IEEEoverridecommandlockouts
\usepackage{cite}
\usepackage{amsmath,amssymb,amsfonts}
\usepackage{algorithmic}
\usepackage{graphicx}
\usepackage{textcomp}
\usepackage{xcolor}
\def\BibTeX{{\rm B\kern-.05em{\sc i\kern-.025em b}\kern-.08em
    T\kern-.1667em\lower.7ex\hbox{E}\kern-.125emX}}
\begin{document}

\title{Collaborative Remote Control of Unmanned Ground Vehicles in Virtual Reality\\
%{\footnotesize \textsuperscript{*}Note: Sub-titles are not captured in Xplore and should not be used}
}

\author{\IEEEauthorblockN{Ziming Li, Yiming Luo, Jialin Wang, Yushan Pan, Lingyun Yu, Hai-Ning Liang\textsuperscript{*}}
\IEEEauthorblockA{\textit{Department of Computing}, \textit{School of Advanced Technology}\\
\textit{Xi’an Jiaotong-Liverpool University}, Suzhou, China \\
\textsuperscript{*} Corresponding author (haining.liang@xjtlu.edu.cn)}
}

\maketitle

\begin{abstract}
Virtual reality (VR) technology is commonly used in entertainment applications; however, it has also been deployed in practical applications in more serious aspects of our lives, such as safety. To support people working in dangerous industries, VR can ensure operators manipulate standardized tasks and work collaboratively to deal with potential risks. Surprisingly, little research has focused on how people can collaboratively work in VR environments. Few studies have paid attention to the cognitive load of operators in their collaborative tasks. Once task demands become complex, many researchers focus on optimizing the design of the interaction interfaces to reduce the cognitive load on the operator. That approach could be of merit; however, it can actually subject operators to a more significant cognitive load and potentially more errors and a failure of collaboration. In this paper, we propose a new collaborative VR system to support two teleoperators working in the VR environment to remote control an uncrewed ground vehicle. We use a compared experiment to evaluate the collaborative VR systems, focusing on the time spent on tasks and the total number of operations. Our results show that the total number of processes and the cognitive load during operations were significantly lower in the two-person group than in the single-person group. Our study sheds light on designing VR systems to support collaborative work with respect to the flow of work of teleoperators instead of simply optimizing the design outcomes.

\end{abstract}


\section{Introduction}
In recent years, a number of studies in the field of computer-supported cooperative work (CSCW) have explored the use of emerging virtual reality (VR) technology as a means of supporting cooperative work \cite{bjorling2022robot}. The development of these cooperative virtual environments (CVEs) represents a radical shift away from the more commonly used single-user VR systems. CVEs are characterized by the rendering of virtual places inhabited by multiple distributed users, who are mutually represented to each other in order to support cooperative action and interaction within the shared virtual space \cite{cruz2021non, chen2021effect, LIANG2019855, chen2020collaborative}. A number of CSCW applications of this technology have previously been reported, including document editing \cite{kovarik2022getting}, video conferencing \cite{han1997cu}, video editing \cite{nguyen2017collavr}, data analysis \cite{cordeil2016immersive}, information gathering \cite{morris2010wesearch} and interacting with virtual objects (building \cite{bjorn2021immersive} and workspace \cite{chow2019challenges}). In addition, a few CSCW studies have addressed the design of remote control centers for autonomous vessels \cite{pan2021reflexivity}. All these mutual reality applications attempt to utilize CVE technology to provide arenas in which social interaction can occur and cooperative work can be supported. 
 
Perhaps surprisingly, many practitioners have focused on VR as a virtual interface for use with CSCW systems \cite{vasarainen2021systematic}. Indeed, for many in the field of CSCW, these applications and related research represent new interface technologies for interacting with computational resources. Thus, they allow for the possibility of using a form of interface to explore the value of social science contributions to the design and evaluation of such systems \cite{ludwig2021share}. For example, in relation to remote control crewless vehicles, prior studies have explored the use of VR to teleoperate robots, including unmanned ground/flying vehicles (UGVs/UFVs) \cite{luo2021device, luo2021Monoscopic}. Such studies have shown that VR effectively supports efficient and accurate teleoperations \cite{bout2017head, islam2019fire}. In fact, the use of a VR interface was found to lower operational costs and reduce the workload in some cases, especially when it comes to UGVs traveling within complex environments \cite{goedicke2018vr, nguyen2019dronevr}. 

However, simply designing CSCW systems for use in VR environments cannot ensure either good cooperation or a balanced workload. As teleoperation becomes more complex, it places a more significant cognitive load on the teleoperator \cite{helsel1992virtual}, which can lead to more operational errors \cite{fakhoury2018effect, sexton2018anticipation, murphy2017load}. Although prior studies have considered familiar control layouts as a means by which interaction designers can reduce the cognitive load facing teleoperators through developing CVE functionalities \cite{nguyen2020user, lee2021xr}, the common failure to properly evaluate the collaboration work that takes place within virtual systems means that it remains questionable whether the cognitive load will actually be reduced. For example, while some studies have investigated the use of VR for learning and education \cite{leung2018use, Lu2018VRTool, pack2020VREAP} and industrial applications \cite{jimeno2016using, Lee2019Tutor}, such as vehicles, cranes and trucks, they have tended to use similar control layouts in an effort to reduce the cognitive load experienced by their operations. These control layouts do not exist in a vacuum. Rather, the design of similar control layouts is in line with the principles of affordance \cite{norman2013design}, articulation work \cite{heath2002configuring, randall2021ethnography} and grounding \cite{clark1991grounding}. For instance, nuclear power plants use standardised information display layouts in order to reduce their controllers’ cognitive load and help their operators to access key information quickly and accurately \cite{clark1991grounding}. Moreover, online learning platforms help students to reduce the cognitive load associated with the learning process by implementing easy and familiar layouts, thereby assisting students in improving their learning efficiency \cite{sweller2019cognitive}. Without understanding the distributed cognition involved in interactive systems, it will be impossible to determine the collaborative work between people within the environment, as well as the resources and materials involved, that can serve to reduce the workload. 

Thus, we consider the cognitive load stemming from cooperative work to represent a mutual means of shaping and reshaping the design of remote control in VR environments. In light of this, our main research question concerns how a collaborative multi-person approach could effectively reduce the cognitive load associated with operating a remote-control vehicle using VR means. More specifically, in the present study, we use VR for the remote control and manipulation of UGVs in order to examine the effectiveness of the task sharing and performance of participants in our pre-designed VR environment. Moreover, we explore whether the remote control of a UGV using a mechanical arm by two participants working collaboratively in a VR interface can be understood as an effective and accurate way of supporting the participants’ collaboration. We conducted two experiments to evaluate our design of collaborative remote control in a VR environment. We simultaneously observed the work needed to complete the required tasks and the total number of operations performed by participants. The findings of our study contribute to the literature by suggesting designs for effective interface layouts for collaborative teleoperations that are relatively complex in nature. 

The remainder of this paper is organized as follows. We present the related work in section \uppercase\expandafter{\romannumeral2}, wherein we position the present work in relation to prior studies. In section \uppercase\expandafter{\romannumeral3}, we discuss the scope of the system and provide an overview to help readers develop a picture of the experimental settings. This is important because it allows for a holistic understanding of how we implemented the technical elements to facilitate collaborative work concerning the remote control of unmanned vehicles in CVEs, as set out in section \uppercase\expandafter{\romannumeral4}. Section \uppercase\expandafter{\romannumeral4} details the experiments, including the applied procedures, formulated hypotheses, and results. Subsequently, section \uppercase\expandafter{\romannumeral6} presents and discusses the findings. Finally, section \uppercase\expandafter{\romannumeral7} concludes the paper by elucidating the limitations of the research.

\section{Related work}
\subsection{Cooperation within virtual environments}
VR, as a virtual interface in CSCW systems, has been the subject of significant research interest in recent years. Such interfaces are designed to support cooperative work by providing a common spatial frame of reference for collaboration of various types and in various contexts. For example, in an organizational setting, VR can take the form of video conferencing or desktop applications, although such applications are not limited to organizational settings or work practices per see and can also provide support for, for example, entertainment \cite{benford1996shared}. Cooperative work occurs in situations where multiple participants are mutually dependent on each other when it comes to completing a shared task and, therefore, are required to engage in articulation work \cite{schmidt2008taking}. This interdependence forms the core of the very definition of cooperative work. To have a cooperative work situation, at least two people must depend on each other to solve a shared task, and in such situations, the need for articulation work arises. In the case of immersive technologies such as VR, augmented reality (AR), and extended reality (XR), articulation work can require that multiple people ‘articulate’ (i.e. divide, allocate, coordinate, schedule, mesh and interrelate) their distributed individual activities through a designed digital object, thereby sharing artifacts and knowledge through a combination of voice, gesture, audio and graphical information that can be used in the real world, virtual world or both to form a flow of work \cite{randall2016common, Lei2021SUI}.

Furthermore, as a virtual interface for cooperative work, VR can be divided into synchronous and asynchronous collaboration with regard to the dynamic changes that occur over time. Remote locations and synchronous collaboration within a virtual environment can support users in working together by immersing them in a co-located setting. In fact, fully immersive virtual environments can support remote and real-time multi-user collaboration, interaction and information and data sharing, which is why they are used in domains such as gaming \cite{brown2004cscw}, entertainment education \cite{pedersen2020virtual}, therapy \cite{begum2015measuring}, training \cite{sun2018design}, decision making and problem solving \cite{morrison2020challenges}. Different tools, features, and functions are available for directly manipulating objects, navigating within the environment, encountering people, and sharing visual artifacts. In addition, collaborative tools and sharing mechanisms such as instant-, audio-, and video messaging can be featured. This enables joint contribution whereby users can simultaneously work together as data are instantly modified. 

When reviewing the prior CSCW research concerning VR, we find that CSCW researchers are aware of how people who engage in work and articulate work use different strategies to reduce the effort associated with articulation work, such as awareness \cite{heath2002configuring} and grounding \cite{clark1991grounding}. When designing VR, we recognize that it is important to imitate real-life work tasks and mimic how cooperative work participants would be allowed to engage in awareness and grounding activities when ‘testing’ the remote control of unmanned vehicles. In most cases, however, the design of awareness cannot be fully understood once the experimental setting is established. Thus, the uncertainty of the VR environment and the limited knowledge of actors’ cooperative work could increase the unnecessary workload among cooperating actors. Most research contexts are taken for granted, and VR practitioners tend to overlook different ways to design for peripheral awareness––allowing participants to ‘see at glance’ who is present and what they do. Therefore, the present study is influenced by previous VR studies but aims to elucidate the capturing and later replaying of multi-modal interactions (i.e. speech, PC, joystick, and scene manipulations) in order to enable cooperating participants to recognize and share when they have a common grounding. This common grounding will help people to expand their grounding references (i.e. pointing to certain artifacts that convey information) and use other cues.

\begin{figure*}[ht]
  \centering
  \includegraphics[width=2.1\columnwidth]{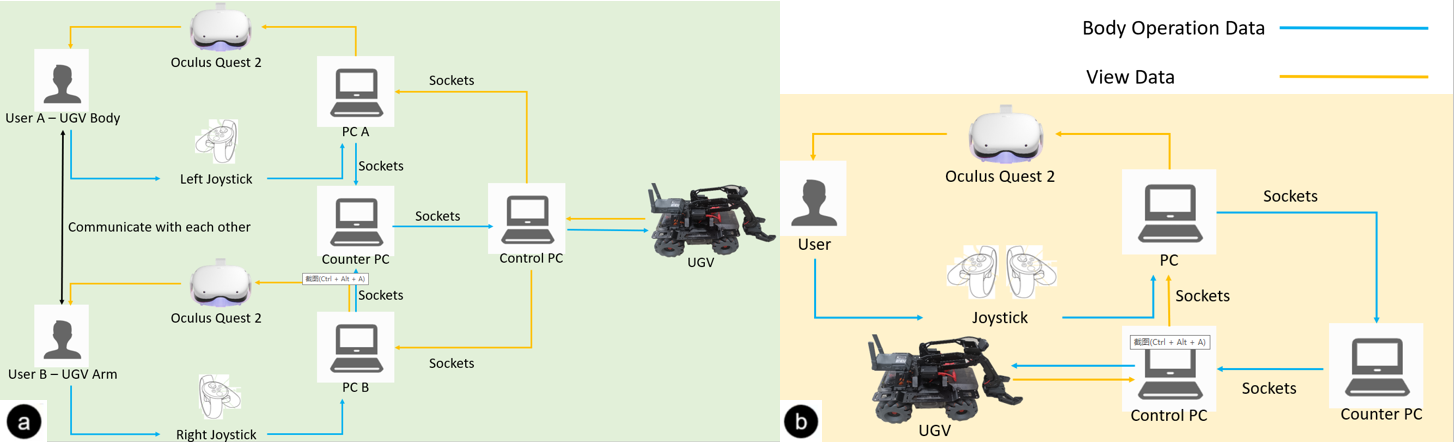}
  \caption{(a: green background) Logical flow chart for the collaboration interaction (Users can communicate with their partner); (b: yellow background) Logical flow chart for the single interaction.
  \label{WholeSys}}
\end{figure*}

\subsection{Measuring the distributed cognitive workload}
Given the above-mentioned issues, we designed a multi-person collaboration experiment intended to measure the cognitive load so as to help us enhance the exploration of the collaborative remote control of UGVs. A reduced cognitive load will allow participants to understand things more quickly and exhibit fewer errors when performing in group setting operations that demand lower levels of cognitive load, for instance, when such operations are divided among multiple users. Yet, the cognitive load has traditionally been associated with individual work \cite{hollan2000distributed, shi2021.PACMGIT, Yu.2019.ToG, Yu.2021.GazeCHI} and used to generate a variety of instructional efforts. Such efforts also influence the effectiveness and efficiency of collaborative work, whether computer-supported or face-to-face work, although they are often not considered when designing CVEs or researching collaborative work practices. The need to consider the cognitive load in collaborative environments involving VR applications is due to the new millennium requiring an enhanced understanding of the emerging dynamics of interaction \cite{roetzel2019information}. The focal task is no longer confined to be desktop but instead reaches into the complex world of information and computer-mediated interactions. VR practitioners are expected to understand the interactions between people and technologies, including the answers to the following questions. How can we coordinate our activities in virtual environments? How do we articulate our work to co-workers? What can we do to ensure that the flow of work in VR is not subject to breakdowns? What techniques or tools can be used to successfully mediate collaborative work? 

One difficulty associated with VR practitioners’ measurement of cognition might be related to the fact that prior human–computer interaction (HCI) research has taken many psychological studies for granted \cite{randall2016common}. Cause-and-effect relationships cannot explain why a high cognitive load can render completing a task difficult \cite{ramakrishnan2021cognitive, stapel2019automated} or why a cooperative workload is sometimes low \cite{costley2018moderating, chu2017effects} and sometimes not \cite{chu2017effects}. This results in a near-endless need for experiments to identify the possible factors that may affect the results (i.e. the p-value) \cite{baker2016statisticians, gelman2016problems}. Designing studies wisely, particularly in the HCI and CSCW fields, is essential \cite{hornbaek2013some}. In line with the approach of previous works, our study focuses on identifying how the way in which participants’ cognitive load is measured during an experiment can significantly impact the outcome of the investigation. The task in question involves two participants working together to complete a remove manipulation and explore its effect on both completion and the perceived workload. Importantly, we do not aim to find a yes or no answer to the question of whether the cognitive load is reduced or not; rather, we hope to identify the cues that enable people to be aware and ground their activities during collaborations so as to trigger a reasonable cognitive workload per person.

\section{System scope and overview}
We developed a VR UGV remote control platform with two versions, one for two participants and another for one, in Unity3D. Therefore, in this study, there were two conditions: (1) Dual-users and (2) Single-users.

\subsection{System architecture and workflow for Collaboration use case}
Figure~\ref{WholeSys}.a shows the system architecture and workflow for our collaboration interaction experiments. There are two users, User A and User B. User A is the source of the UGV’s body movement signal in the system via a left joystick. User B is the source of the UGV’s arm control signal in the system via a right joystick. Both users can receive the version data of the UGV through the Oculus Quest 2’s HMD. The principle is that the Control PC must receive the camera singles from the RoboMaster before being simultaneously distributed to PC A and B. This step is vital because, without such singles, it is impossible to create immersive environments for both participants to control the UGV.

In addition to the intervention of the two Oculus Quest 2 sets, the collaborative experiment was completed with the participation of these devices: two laptops that support Oculus Link (a data cable used to help Oculus Quest 2 HMD rendering with the help of the computer's rendering performance), a computer that was used to assist in transmitting and counting the operational commands used and a computer that could control the UGV over the LAN.

In the collaborative experiment, there are two types of signals. One is the control signal for the movement of the vehicle's body, which is sent by User A in Figure 1 with the help of an Oculus Quest 2 left-handed joystick and is transmitted to PC A. PC A receives the signal for the movement of the vehicle and sends it directly to the Counter PC via the LAN. Similarly, the signal from User B to control the arm is sent from PC B to Counter PC by the same method with the help of a right-handed joystick. Once the Counter PC has received both signals, it will count them and send them to the Control PC controlling the UGV through LAN. Finally, the UGV will operate on the basis of the signals received.

The onboard lens from the UGV sends its field of view data in real-time to Control PC, which sends it via LAN to PC A and PC B. After that, PC A and PC B will help the Oculus HMD to display the field of view data and present it to the users. We use a Counter PC to facilitate singles before sending them to the Control PC for directly outputting all experiment data as a whole. This helps to reduce the time complexity for outputting each experiment data. In that case, data from PC A and PC B can be simultaneously transformed to the Control PC without disruption, neither from PC A nor PC B. 

\subsection{System architecture and workflow for single use case}
Figure~\ref{WholeSys}.b shows the system architecture and workflow for our single interaction experiments. Unlike collaborative experiments, single experiments will have one user doing all the controlling of the UGV. UGV's view data will also only be presented to this user.

The data transfer method for the single experiment is the same as for the collaborative experiment, with the only difference being that each group uses one less Oculus Quest 2 and one less laptop to assist Oculus with rendering and relaying the data.

\section{Experiment}
\subsection{User arrangement}
A total of 12 participants (4 females and 8 males, aged between 20-23, M = 21.5) were invited to join this study and divided into 8 groups according to the between-subjects experiment design. 4 pairs (8 participants) randomly formed were assigned to the two-person collaborative group. The other four participants were assigned to the single-user group. Data collected from the pre-experimental questionnaire showed that all participants had no previous experience remotely controlling a UGV via a VR interface. They did not report any physical symptoms and had normal or corrected-to-normal vision. Prior to the actual experiment, they were given a pre-training practice to allow them to become familiar with the VR device, controls, and remote manipulation. All participants successfully completed the pre-training.

\begin{figure*}[t]
  \centering
  \includegraphics[width=2.0\columnwidth]{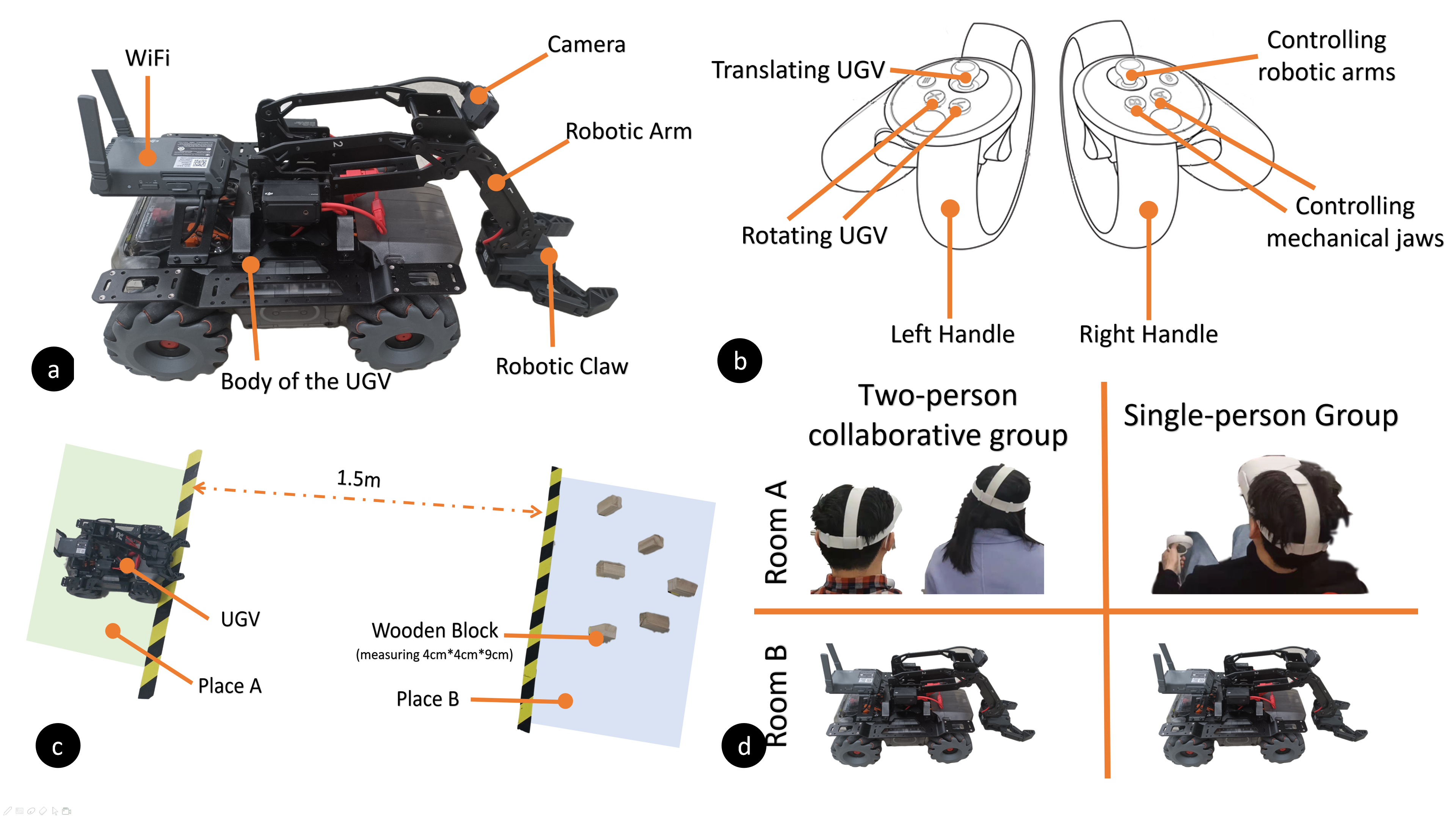}
  \caption{(a) Structure of the UGV used in the study; (b) Description of how participants control the UGV using the Oculus Touch controllers; (c) A picture of the experimental task; and (d) The experiment environment design.
  \label{TotalFigure}}
\end{figure*}

\subsection{Devices Used}
For this study, we have used the RoboMaster EP robot~\footnote{https://www.dji.com/hk/robomaster-ep}. The body of this robot can be divided into a vehicle body, a robotic arm, a camera, and a WiFi module (as shown in Fig.~\ref{TotalFigure}a). Participants were able to observe the field of view from the robot's camera through the Oculus Quest 2 and use the Oculus Touch controller (see Fig.~\ref{TotalFigure}b\footnote{https://docs.unity3d.com/560/Documentation/Manual/OculusControllers.html}.) to operate the robot's body movement and control the robotic arm.

Participants could control the UGV'S translation and rotation using the left handle (see Fig.~\ref{TotalFigure}b). Participants could also use the right handle to control the robotic arms and jaws (see Fig.~\ref{TotalFigure}b).

\subsection{Task and Procedure}
We asked participants to move the blocks (each measuring 4cm$\times$4cm$\times$9cm) from one place to another using the RoboMaster EP (see Fig.~\ref{TotalFigure}c). This task of moving the blocks was designed to allow the participants to control the robot's arm and move it simultaneously. As shown in a pilot study, the dual nature of the task was challenging enough to elicit a desired level of cognitive load. 

Both single-person and two-person groups had to wear the Oculus Rift 2 and use the VR remote control platform we implemented and run in Unity to control the UGV remotely. In the single-person group, participants were required to operate the movement of the UGV and control the robotic arm. In the two-person collaborative group, one participant needed to control the movement of the UGV and the other participant had to control the robotic arm. During the experiment, the UGV and the participants were in two different rooms to realistically simulate the remote control of the UGV (see Fig.~\ref{TotalFigure}d). Although the participants were seated in the same room and had the opportunity to orally communicate directly, our purpose was to provide them with visual cues in the collaborative immersive environments to initiate collaboration. Thus, our experiment paid attention to the designed tasks considering direct communication as a default condition.  

A simple pre-training was given to participants to complete before the formal experiment. The purpose of this pre-training was to enable participants to become familiar with remote control operation of the UGV in a VR environment, and also to enable participants in the two-person group to become familiar with collaborative remote control operations. In the pre-training participants had to try to pick up a wooden block using the UGV and move it to a designated place. Immediately after the pre-training, participants started the formal experiment. 

In the formal experiment, participants were given the following five tasks to complete: (1) Pick up wooden blocks (6 in total) from one side of a line (Place B); (2) Rotate the UGV and orient it towards the other side of the line (Place A); (3) Move the UGV to Place A and put the block down; (4) If there were still blocks in Place B, move the UGV to Place B and pick up these blocks; (5) Repeat tasks 1 to 4 until there are no more blocks in Place B. Upon completion of the experiment, participants were asked to complete the NASA-TLX questionnaire to measure their cognitive load during the experiment. Participants also needed to complete a Semi-Structured user experience questionnaire (UEQ) (7-point scoring scale) to allow us collect participants' ratings of the operational difficulty of the remote control approach used in the experiment.

\subsection{Hypotheses}
To answer our research question, we formulated the following four hypotheses. We focus on two particular points to investigate how articulation work is supported in VR-based collaboration: (1) the VR interface itself, and (2) social interaction in collaborative interaction in the VR environments, such as awareness and communication. Additionally, those artifacts which were used to coordinate cooperative interaction can also be a vital factor to enable the collaborative remote control of unmanned vehicles. 
\begin{enumerate}
	\item[$\bullet$] $H_1:$ with the help of the other participant, participants in the collaborative group would take less time to complete the tasks than those in the single-user group.
	\item[$\bullet$] $H_2:$ based on Brook's Law, participants in the two-person group may use more actions to complete a task than participants in the single-person group because of the higher level of discussion and actions needed during the collaboration process (that is, participants in this group could make more missteps).
	\item[$\bullet$] $H_3:$ with the assistance of another participant, participants in the two-person group would experience a lower cognitive load in completing the experiment.
	\item[$\bullet$] $H_4:$ compared to the single-person group, the two-person group would have a lower level of difficulty to learn the operation of control UGV.
	\item[$\bullet$] $H_5:$ compared to the single-person group, the two-person group would have a lower level of difficulty to complete the given task.
\end{enumerate}

\section{Results \& analysis}
All participants in this experiment understood and completed the given experimental tasks successfully. The data we collected were valid. For the objective data we collected during the experiment, the results of the Shapiro-Wilk test show that our data were normally distributed ($p > .05$). Results of Levene's test show that the assumption of homogeneity of variance was met in each condition ($p > .05$). We conducted independent t-tests for the comparison of objective measurements. Similar to the objective data, the subjective data also return the same results from Levene's test ($p > .05$). Thus, we conducted the Mann-Whitney U test for the pairwise comparison of subjective measurements.

\subsection{Objective Results}
The time spent by participants to complete the task was obtained directly from the length of the video recorded during the experiment. The number of operations performed by the participants to complete the task was counted directly by the Unity application. As many of the commands issued by participants during the remote control process, such as UGV forward and backward, were continuous commands, the application treated and counted the commands as a state. It counted each time the participant switched between states.

\begin{figure}[htp]
  \centering
  \includegraphics[width=1.0\columnwidth]{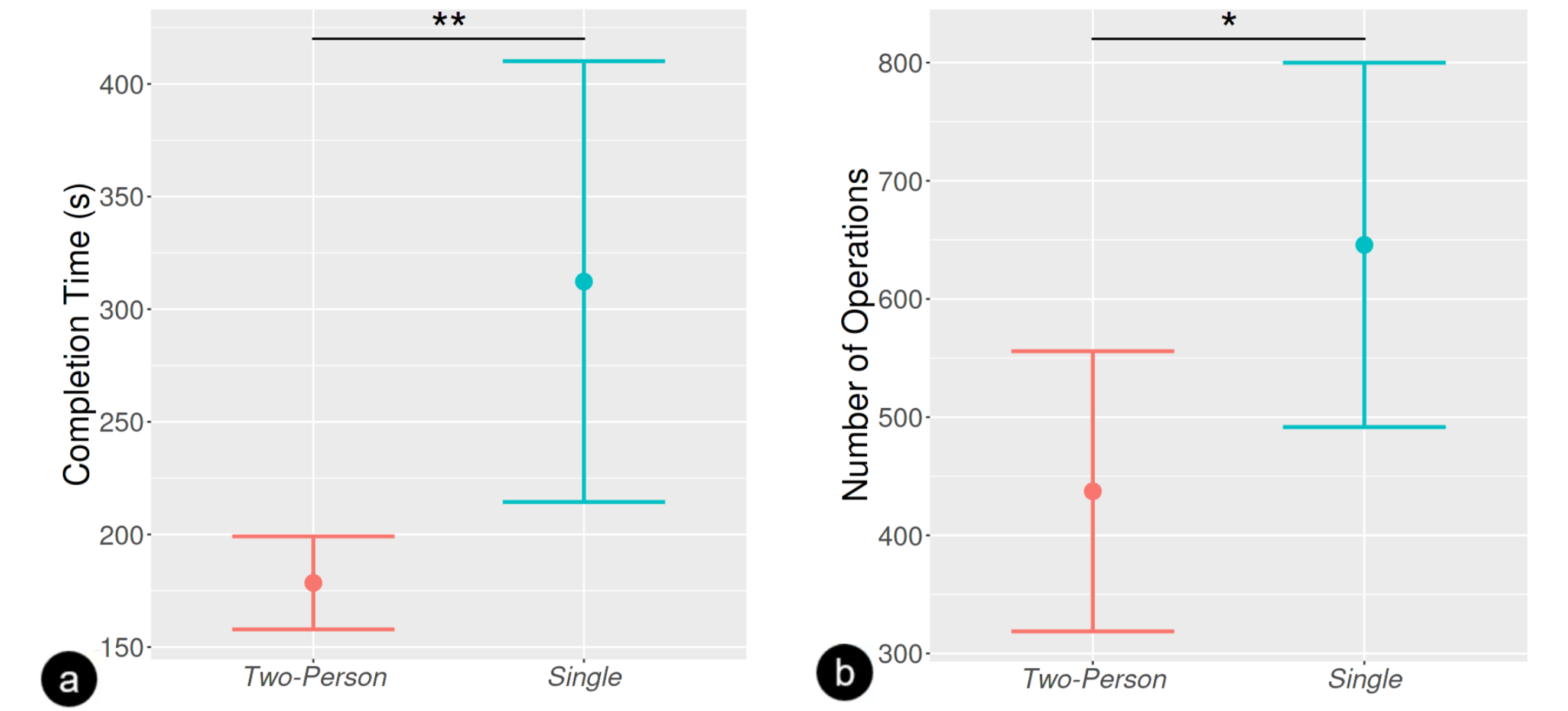}
  \caption{(a) Mean completion time(s) of the task, and (b) mean number of operations. The error bars represent 95\% confidence intervals. '**', '*' represent a '.01' and '.05' significance levels, respectively. Two-person: Two-person collaborative group; Single: Single-person group \label{objdata}}
\end{figure}

Independent t-test shows that the mean completion time (s) of the task differed significantly between two groups ($t(6) = -4.256, p < .01$). Fig.~\ref{objdata}a shows the mean completion time of the two-person collaborative group and single-person group. It shows that the two-person group took much less time to complete the experimental task than the single-person group, which supports $H_1$.

The two-person collaborative group took less time to complete tasks than the single-person group for two reasons. One is because each participant in the two-person group needed to control part of the UGV, and therefore this  control was less difficult for these participants than for the single-person group. The other reason is that participants in the two-person group had less information to focus on during task completion, i.e. the cognitive load was less. Both reasons were confirmed in the subsequent analysis of the subjective data from the questionnaire.

Results of t-tests on the mean number of operations also shows a significantly difference between two groups ($t(6) = -3.412, p < .05$). Fig.~\ref{objdata}b shows the mean number of operations of each group. It shows that the two-person group used significantly fewer operations to complete the experimental task, which does not support $H_2$, to our surprise. From our post-experiment interviews with participants, we found the main reason why the two-person group completed fewer total actions than the single-user group. Although there were some incorrect operations due to disagreements between some pairs in the two-person group, there were also incorrect actions made by single users due to cognitive overload. As such, we can infer that more errors were made in the single group for a different reason than in the two-person group, resulting in $H_2$ not being supported.

%Based on our observations, it seems that participants in the two-person group may regularly behave independently, without discussing much with their teammate.  
%misbehave because they have different goals than their teammates. However, the results of the analysis of this experiment prove that our conjecture is not valid. In post-experimental interviews, we learned that although participants in the two-person group did misbehave because they did not match their teammates' goals, the number of misbehaviour in the two-person group was lower than the number of misbehaviour in the single-person group due to cognitive overload.
% the misbehaviour in the two-person group was lower than the number of  misbehaviour in the single*-person group might not because of cognitive overload. It might cause by distributed cognitive/awareness/articulation(if they talk to each other), or the VR itself enables a workflow of articulation work to reduce anxious of work. That might reduce the cognitive overload. But I would like to ask if you have more data to support this finding?

\subsection{Subjective Results}
The results of the Mann-Whitney U test for NASA-TLX questionnaire data show that the cognitive load of the two-person  group was significantly lower than the single-user group ($p < .05$). This result supports $H_3$. As for the UEQ data's testing, the test results for UEQ data did not show significant differences in the learning difficulty between the two groups ($p > .05$), which does not support $H_4$. However, the test results for UEQ data showed that there were significant differences in perceived difficulty of operating the UGV between the two groups ($U = 2.5, p =.019.05$). The results of the descriptive analysis of the UEQ data show that participants in the two-person group perceived the operational difficulty of the remote control of the UGV using the VR interface to be low, whereas participants in the one-person group perceived the difficulty to be medium to high. This result supports $H_5$. This suggests that the use of two-person control significantly reduces the operational difficulty when controlling a UGV remotely using VR.

Based on these results, we found that two-person collaboration significantly helped participants to reduce the operational difficulty. It also significantly reduced the cognitive load of participants when performing the task.

\section{Discussion}
From the above analysis, we can see that, when compared with the single group, the collaborative group took significantly less time to complete the experimental task (supports $H_1$); the total number of operations to complete the experimental task was lower (supports $H_2$); the cognitive load was significantly less (supports $H_3$); there was no significant difference in the difficulty perceived by the volunteers when learning to operate the UGV (supports $H_4$), and the difficulty perceived by the volunteers when operating the UGV during the experiment was significantly lower (supports $H_5$). These outcomes are attributable not only to the collaborative versus single group dichotomy but also to additional characteristics of the structure of the UGV remote control system that we designed in this project.

In the framework of our system, participants in collaborative groups can communicate their task goals directly with their partners without delay. The establishment of $H_1$ and the non-establishment of $H_2$ is strongly related to this feature of the experimental system. This is because participants in the collaborative group can confirm the target with their partner easily and quickly via direct communication, so participants in the collaborative group expended little time and manipulation in confirming the common target. We also realized that the experiment's sample size was relatively small, and the population might not be representative. These are biases of the investigation. Due to Covid-19, it was hard to recruit a significant sample size with a wide age range. If more participants in the collaborative group and their partners were not in the same room when the experiment was conducted or could not communicate with each other, the results of the experiment for $H_1$ and $H_2$ might have been judged differently. However, our designed system fruitfully confirmed the articulation work of teleoperators' remote control of unmanned ground vehicles. This articulation of collaboration is essential, and it opens room for other researchers to explore systems development for collaboration in immersive environments.

To generate additional supporting evidence, the design of the experimental system for remote control operation separates the operation of the different functional areas of the UGV into different handles. This feature is manifested in the fact that the control of the movement of the vehicle is executed entirely by the left joystick and the control of the robotic arm is executed entirely by the right joystick. This design reduces the learning difficulty for participants and leads to the non-establishment of $H_4$. This design also makes it less likely that participants will confuse the operation when they only need to control one function, which validates $H_3$ and $H_5$.

Finally, the collaborative characteristics of the designed system are also supported by the final results of the experiment. The participants invited into this experiment had never had any experience of using a joystick to remotely control a UGV. These participants also did not have experience with remote control of UGVs. The experiment results show that our design helps teleoperators access key information quickly and accurately. In turn, their cognitive loads were reduced when using easy user interfaces.

\section{Conclusion}
This study built a remote control platform for UGVs in a VR environment that supports two-person collaboration. We then used this platform to evaluate how users' perceptions, work states, and cognitive load differ when remotely controlling UGVs in this platform compared to traditional single-person remote control UGVs. We illustrate the scope of effectiveness and conditions of applicability of collaborative control in UGV remote control in VR environments. Our system shed light on the importance of awareness of multi-person collaboration and the articulation work in cooperative remote control of UGVs in VR environments. We assert that using VR as an interface for supporting remote control shall enhance the collaboration among multi-users. This contribution opens a pathway to focus on identifying and investigating those cues, tools, and coordination mechanisms to maintain the flow of work in VR environments. VR-supported cooperative systems, with such focal points as remote control, multiple teleoperators, and cognitive workload are fruitful areas for further study. Collaborative work could be greatly enhanced with further development of collaborative remote control in the VR field.

\section*{Acknowledgement}
We thank the participants who generously shared their time to do the experiment.  This work was supported in part by Xi’an Jiaotong-Liverpool University (XJTLU) Key Program Special Fund (\#KSF-A-03) and XJTLU Research Development Fund (\#RDF-17-01-54, \#RDF-21-02-008, and \#RDF-19-02-11).

\bibliographystyle{IEEEtran}
\bibliography{refs}

\end{document}